
\input harvmac
%
%
%
%
\ifx\answ\bigans
\else
\output={
  \almostshipout{\leftline{\vbox{\pagebody\makefootline}}}\advancepageno
}
\fi
%
%
\def\mayer{\vbox{\sl\centerline{Department of Physics 0319}
\centerline{University of California, San Diego}
\centerline{9500 Gilman Drive}
\centerline{La Jolla, CA 92093-0319}}}
%
%
\def\doe{\#DOE-FG03-90ER40546}
\def\tnlrc{\#RGFY93-206}
%
%
\def\UCSD#1#2{\noindent#1\hfill #2%
\bigskip\supereject\global\hsize=\hsbody%
\footline={\hss\tenrm\folio\hss}}
%
%
\def\abstract#1{\centerline{\bf Abstract}\nobreak\medskip\nobreak\par #1}
%
%
%
%
\edef\tfontsize{ scaled\magstep3}
 \tfontsize  \tfontsize
 \tfontsize \font\titlei=cmmi10 \tfontsize
\font\titleis=cmmi7 \tfontsize \font\titleiss=cmmi5 \tfontsize
\font\titlesy=cmsy10 \tfontsize \font\titlesys=cmsy7 \tfontsize
\font\titlesyss=cmsy5 \tfontsize  \tfontsize
\skewchar\titlei='177 \skewchar\titleis='177 \skewchar\titleiss='177
\skewchar\titlesy='60 \skewchar\titlesys='60 \skewchar\titlesyss='60
%
%
%
%
\def\inv{^{\raise.15ex\hbox{${\scriptscriptstyle -}$}\kern-.05em 1}}
\def\lbar{{\lower.35ex\hbox{$\mathchar'26$}\mkern-10mu\lambda}} 

\def\kahler#1#2{K_{#1 {#2^*}}}
%
%
\def\dsl{\,\raise.15ex\hbox{/}\mkern-13.5mu D} 
\def\delsl{\raise.15ex\hbox{/}\kern-.57em\partial}
\def\Ksl{\hbox{/\kern-.6000em\rm K}}
\def\Asl{\hbox{/\kern-.6500em \rm A}}
\def\Dsl{\hbox{/\kern-.6000em\rm D}} 
\def\Qsl{\hbox{/\kern-.6000em\rm Q}}
\def\gradsl{\hbox{/\kern-.6500em$\nabla$}}
%
%
\def\lspace{\ifx\answ\bigans{}\else\qquad\fi}
\def\lbspace{\ifx\answ\bigans{}\else\hskip-.2in\fi} 
%
%
\def\boxeqn#1{\vcenter{\vbox{\hrule\hbox{\vrule\kern3pt\vbox{\kern3pt
        \hbox{${\displaystyle #1}$}\kern3pt}\kern3pt\vrule}\hrule}}}
%
%
\def\mbox#1#2{\vcenter{\hrule \hbox{\vrule height#2in
\kern#1in \vrule} \hrule}}
%
%

  \def\CO{{\cal O}}

%
%

\def\partder#1#2{{\partial #1\over\partial #2}}

\def\bar#1{\overline{#1}}
\def\vev#1{\left\langle #1 \right\rangle}

\def\abs#1{\left| #1\right|}

\def\darr#1{\raise1.5ex\hbox{$\leftrightarrow$}\mkern-16.5mu #1}

%
%
\def\frac#1#2{{\textstyle{#1\over #2}}}
\def\[{\left[}
\def\]{\right]}
\def\({\left(}
\def\){\right)}
%
%

\def\weff{{W_{\rm eff}}}
%
%
\def\ltap{\ \raise.3ex\hbox{$<$\kern-.75em\lower1ex\hbox{$\sim$}}\ }
\def\gtap{\ \raise.3ex\hbox{$>$\kern-.75em\lower1ex\hbox{$\sim$}}\ }
\def\gl{\ \raise.5ex\hbox{$>$}\kern-.8em\lower.5ex\hbox{$<$}\ }
\def\roughly#1{\raise.3ex\hbox{$#1$\kern-.75em\lower1ex\hbox{$\sim$}}}
%
\def\ie{\hbox{\it i.e.}}        
\def\eg{\hbox{\it e.g.}}        
\def\etal{\hbox{\it et al.}}

\def\np#1#2#3{Nucl. Phys. B{#1} (#2) #3}
\def\pl#1#2#3{Phys. Lett. {#1}B (#2) #3}
\def\prl#1#2#3{Phys. Rev. Lett. {#1} (#2) #3}
\def\physrev#1#2#3{Phys. Rev. {D#1} (#2) #3}

\def\prep#1#2#3{Phys. Rep. {#1} (#2) #3}

\def\cmp#1#2#3{Comm. Math. Phys. {#1} (#2) #3}

\def\lfm#1{\medskip\noindent\item{#1}}
\def\susy{supersymmetry}
\def\Susy{Supersymmetry}
\relax
\noblackbox
\centerline{{\titlefont{R Symmetry Breaking Versus}}}
\medskip
\centerline{{\titlefont{\Susy\  Breaking}}}
\vfill
\centerline{Ann E. Nelson }
\medskip
\mayer
\vfill
\centerline{and}
\centerline{Nathan Seiberg }
\medskip
\vbox{\sl\centerline{Department of Physics and Astronomy}
\centerline{\sl Rutgers University}
\centerline{\sl Piscataway, NJ 08855-0849}}
\vfill
\abstract{
We point out a connection between R symmetry and \susy\ breaking.
We show that the existence of an R symmetry is a necessary condition for
\susy\ breaking and a spontaneously broken R symmetry is a sufficient
condition provided two conditions are satisfied.  These conditions are:
{\it genericity}, \ie\ the effective Lagrangian is a generic Lagrangian
consistent with the symmetries of the theory (no fine tuning), and {\it
calculability}, \ie\ the low energy theory can be described by a
supersymmetric Wess-Zumino effective Lagrangian without  gauge
fields.  All known models of dynamical supersymmetry breaking possess
such a spontaneously broken R symmetry and therefore contain a
potentially troublesome axion. However, we use the fact that genericity
is {\it not} a feature of supersymmetric theories, even when
nonperturbative renormalization is included, to show that the R symmetry
can in many cases be explicitly broken without restoring supersymmetry
and so the axion can be given an acceptably large mass.}

\vfill
\UCSD{\vbox{\baselineskip=9pt\hbox{UCSD/PTH 93-27}\hbox{RU-93-42}
\hbox{hep-ph@xxx/9309299}}}{September 1993}

\newsec{Introduction}
Dynamical \Susy\ Breaking (DSB) provides an attractive way
to achieve a large hierarchy between the Planck scale $M_P$ and the scale
of \susy\ breaking, and, in models where the weak scale is related to
the \susy\ breaking scale, can solve the gauge hierarchy problem
\ref\witten{E. Witten, \np{188}{1981}{513}}.
In DSB models, all mass scales, including the scale of \susy\ breaking,
arise through dimensional transmutation, and thus are proportional to
$M_P e^{-a/g^2}$, where $a$ is a constant of order $4\pi^2$ and $g$ is
some effective coupling constant at the Planck scale. There are several
examples of models which exhibit DSB
\nref\adstwo{I. Affleck, M. Dine and N. Seiberg, \pl{137}{1984}{187};
\prl{52}{1984}{493}; \pl{140}{1984}{59}; \np{256}{1985}{557}}
\nref\otherdsb{G.C. Rossi and G. Veneziano, \pl{138}{1984}{195};  Y.
Meurice and G. Veneziano, \pl{141}{1984}{69}; D. Amati, G.C. Rossi and G.
Veneziano, \pl{138}{1984}{195}; D. Amati, K. Konishi, Y. Meurice, G.C.
Rossi and G. Veneziano, \prep{162}{1988}{169}}
\refs{\adstwo,\ \otherdsb}.  In reference \adstwo\ the following
guidelines for finding dynamical \susy\ breaking models were suggested:
\lfm{1.} The classical potential should not have any noncompact flat
directions, \ie\ it should be nonzero for large field strengths in all
directions in field space.
\lfm{2.} The theory should contain a nonanomalous continuous global
symmetry, and the complexity or absence of solutions to 't Hooft's
anomaly matching conditions
\ref\thooft{G. 't Hooft, in {\it Recent Developments in Gauge Theories},
ed. G. 't Hooft \etal, (Plenum, New York, 1980)}
should imply that this symmetry is spontaneously broken.
\medskip

These conditions are generally sufficient for spontaneous \susy\
breaking because if \susy\ is unbroken, then the Goldstone boson {}from
the spontaneously broken global symmetry should reside in a
supermultiplet which also contains a massless fermion and another
massless scalar. In a few cases, this second massless scalar could
itself be a Goldstone boson of another spontaneously broken symmetry, or
even if it is not a Goldstone boson, its target space can be compact.
Otherwise the existence of another massless scalar implies an exact,
non-compact vacuum degeneracy (flat directions).  However the first
condition is that classically there is no such degeneracy. For a
nonperturbatively generated term to restore the degeneracy which is
absent classically is implausible and usually not possible since in
asymptotically free theories the dynamically generated terms can be
shown to be less important for large field strengths than the classical
terms.

No argument has been given that these criteria are {\it necessary} for
DSB, yet all known examples satisfy them by having no flat directions
and a spontaneously broken $U(1)$ R symmetry.  An argument based on the
anomaly
\ref\kon{K. Konishi, \pl{135}{1984}{439}}
\eqn\konishi{\{\bar Q_{\dot \alpha}, \psi^{\dagger \dot
\alpha}_i \phi_i\}= \partial_i W \phi^i + C {g^2 \over 32 \pi^2}
\lambda\lambda}
($W$ is the superpotential, $\lambda$ is a gluino and $C$ is a constant
depending on the gauge group) was given in reference \adstwo\ suggesting
that in many gauge theories an exact, spontaneously broken R symmetry is
a {\it sufficient} condition for \susy\ breaking. In section~2 of this
paper we argue that such an R symmetry is a feature of any DSB model
where the low energy effective theory can be studied by integrating out
the gauge dynamics, and in which the low energy effective Lagrangian is
a generic Lagrangian consistent with the symmetries (no fine tuning). The
existence of a spontaneously broken $U(1)$ R symmetry is therefore a
useful guideline for model builders attempting to break \susy.

This line of reasoning leads to a serious problem.  Spontaneously broken
R symmetry implies the existence of an exactly massless Goldstone boson,
the {\it R-axion}\foot{We refer to this Goldstone boson as an axion even
though it is exactly massless.}.  General theoretical arguments based on
quantum gravitational effects
\ref\worm{S. Hawking, \cmp{43}{1975}{199}; \pl{195}{1987}{337}, G.V.
Lavrelashvili, V. Rubakov, and P. Tinyakov, JETP Lett. 46 (1987)167; S.
Giddings and A. Strominger, \np{307}{1988}{854}; S. Coleman,
\np{310}{1988}{643}; T. Banks, Physicalia, 12 (1990) 19},
and on string theory
\ref\noglob{T. Banks and L. J. Dixon, \np{307}{1988}{93} },
lead us to expect that any apparent global symmetries in an effective
Lagrangian are accidental consequences of gauge symmetries and
renormalizability, and are only approximate. (R symmetry is difficult to
gauge because of the necessity of anomaly cancellation.)  R symmetry
breaking above the \susy\ breaking scale is also necessary in order to
be able to tune the cosmological constant to zero.  Furthermore, there
are strong phenomenological constraints on an exactly massless Goldstone
boson.  We conclude that to describe our world, any continuous R
symmetry is likely to be explicitly broken.  However, if an R symmetry
is a necessary condition for \susy\ breaking, and Nature does not have
such an R symmetry, \susy\ cannot be spontaneously broken!

In section~3 we examine the effects of explicit R symmetry breaking in
the effective Lagrangian, in order to see whether \susy\ can be broken
without an exact R symmetry. We find that there are many cases where one
can show that the effective Lagrangian is {\it not} the most general
consistent with the symmetries, even when nonperturbative effects are
considered, and that \susy\ {\it is} spontaneously broken without an
exact R symmetry.

\newsec{Criteria for \susy\ breaking}

In flat space, assuming there are no Fayet Iliopoulos terms in the
Lagrangian\foot{since we are only interested in models where all mass
scales arise dynamically }, \susy\ is unbroken at tree level if and only
if there exist some values of the scalar components of the chiral
superfields $z_i$ for which
\eqn\unbroken{\eqalign{\partder{ }{z_j} W&=0\quad ({\rm ``F\ terms"})\cr
z^*_j T^A_{jk} z_k&=0\quad ({\rm ``D\ terms"})\ , \cr }}
where $W$ is the superpotential, the Kahler potential $K$ is $\sum_j
z^*_jz_j$, and the $T^A$'s are the gauge generators. Since in general
these are $m$ complex conditions and $r$ real conditions for $m$
complex unknowns, where $m$ is the number of chiral superfield components
and $r$ is the number of gauge generators, one might naively suppose that
\susy\ breaking would be easy. In fact spontaneous \susy\ breaking is
surprisingly difficult to achieve
\ref\wittentwo{E. Witten, \np{202}{1982}{253}}.

Many interesting examples of gauge theories are known where effects of
the gauge dynamics may be systematically computed.  In such theories
there is a limit where a superpotential coupling can be taken to zero,
making the vacuum energy, and thus the \susy\ breaking scale,
arbitrarily small. Thus the scale of \susy\ breaking is lower than the
scale at which the gauge dynamics becomes strong.  Then the gauge
dynamics may be integrated out and an effective supersymmetric theory
which contains only the light chiral superfields can be constructed. The
following argument shows that in many such theories a supersymmetric
vacuum is absent if there is a spontaneously broken continuous R
symmetry.

We first construct an effective Lagrangian for the light degrees of
freedom, in the limit where the couplings in the classical
superpotential are weak.  In general there are solutions to the D-term
equations in eq.~\unbroken, which depend on a finite number of
continuous parameters (``D-flat directions''). These represent the
massless scalar degrees of freedom in the classical limit with no
superpotential. For some theories one can find D-flat directions where
the gauge symmetry is completely broken and the low energy effective
theory has no gauge fields \adstwo. In other theories there are D-flat
directions where the gauge symmetry is not completely broken, but all
chiral superfields which transform nontrivially under the unbroken
subgroup get large masses, and the remaining low energy effective theory
consists of light gauge singlet fields and a supersymmetric gauge
theory, but with no light chiral superfields carrying the unbroken gauge
symmetries \adstwo.  In both cases the effects of the nonperturbative
gauge dynamics are understood using various methods
\nref\nonpert{G.  Veneziano and S. Yankielowicz, \pl{113}{1982}{321};
T.R. Taylor, G. Veneziano, and S. Yankielowicz, \np{218}{1983}{493};
V.A. Novikov, M.A. Shifman, A. I.  Vainshtain and V. I. Zakharov,
\np{223}{1983}{445}; M. Peskin, in {\it Problems in Unification and
Supergravity}, ed. G. Farrar and F. Henyey (AIP, New York, 1984)}
\nref\adsone{I. Affleck, M. Dine, and N. Seiberg, \prl{51}{1983}{1026};
\np{241}{1984}{493}}
\refs{\adstwo,\ \wittentwo,\ \nonpert,\ \adsone} and the scale of \susy\
breaking can be made arbitrarily small compared to the scale at which
the gauge dynamics become strong by tuning the superpotential
parameters.

We will refer to either of these cases as {\it calculable} theories.
The light chiral superfields $\phi_i$ may be constructed as gauge
invariant functions of the parameters describing the D-flat directions.
The effective supersymmetric Lagrangian has no gauge terms, but
typically has a very complicated Kahler potential $K(\phi_i,
\phi_i^\dagger)$.

We now write the superpotential $W(\phi_i)$ and make the crucial
assumption of {\it genericity}, \ie\ that the gauge dynamics will
generate nonperturbatively additional terms in $\weff$ which are generic
(locally) holomorphic functions of the fields $\phi_i$ consistent with
the global symmetries of the theory.  In other words, the low energy
effective Lagrangian has no fine tuned coefficients.  This assumption of
genericity is known to be satisfied in standard field theories.  It has
been shown explicitly to be true in supersymmetric theories for a number
of examples (see, \eg\ ref. \refs{\adstwo,\ \adsone}), but there are
also cases where it is {\it not} true
\ref\nonren{N. Seiberg, to appear}.

If we restrict ourselves to values of $\phi$ for which $\kahler ij$
is finite and nonsingular (which typically means $\phi_i \not= 0,
\infty$), the criterion for unbroken \susy\ in the effective theory is
\eqn\fterm{\partder{\weff}{\phi_i} =0\ .}

We now examine whether or not there is a supersymmetric minimum for
various possibilities for the global symmetries:
\lfm{\it 1.}{\it No symmetries.} If there are no global symmetries the
conditions~\fterm\ represent $n$ equations for $n$ unknowns, which are
soluble for generic $\weff$, hence \susy\ should be unbroken. It is
possible to find special superpotentials with no global symmetries for
which the conditions \fterm\ are not soluble, however these do not
satisfy the genericity assumption.  They are unstable under small
variations in the couplings.
\lfm{2.}{\it Continuous global symmetry which commutes with \susy.}
Constraining the theory by means of a non-R global symmetry does not
allow one to evade the conclusion that \susy\ is unbroken. Because
$\weff$ is holomorphic, if there are $l$ generators of global
symmetries, $\weff$ may be written as a function of $n-l$ variables. For
example, in the case of a $U(1)$ global charge where the fields $\phi_i$
have charges $q_i$, $\weff$ is a function of the $n-1$ variables
$X_i=\phi_i/{\phi_n^{q_i/q_n}}$, where $i=1,\ldots,n-1$, and
$q_n\not=0$. Thus since $\weff$ is independent of $l$ complex variables,
$l$ of the equations \fterm\ are automatically satisfied.  The other
equations give $n-l$ constraints for $n-l$ unknowns, which typically
have solutions.  The counting of equations and unknowns in the case of a
global discrete symmetry is as in case 1 above (no symmetries).
\lfm{3.}{\it Spontaneously broken $U(1)$ R symmetry.} The superpotential
carries charge 2 under an R symmetry, and so if $\phi_n$ receives an
expectation value and carries  R charge $q_n$, $\weff$ may be written
\eqn\weqn{\weff=\phi_n^{2/q_n}f\(X_i\)\quad ; \quad
X_i={\phi_i\over{\phi_n^{q_i/q_n}}},}
where $i<n$ and $f$ is a generic function of the $n-1$ variables $X_i$.
Now for \susy\ to be unbroken, the $n$ equations
\eqn\nmo{\partder{f}{X_i}=0}
and
\eqn\nth{f=0}
must be satisfied.  (In deriving \nmo\ and \nth\ we used the fact that
$\phi_n$ is finite and nonzero.)  These are $n$ equations with $n-1$
unknowns which cannot be satisfied for generic $f$.  Therefore,
generically there is no supersymmetric solution. (It is however possible
to find special superpotentials which will spontaneously break R but not
\susy; it is implausible that such a superpotential could be generated
dynamically.) One may worry that there could be a supersymmetric minimum
in the limit where the   R symmetry is restored. Because the change of
variables  in \weqn\ is singular in this limit, it is
necessary to examine the behavior of ${\kahler ij}^{-1}$
to decide whether or not \susy\ is restored as $\phi_n\rightarrow0$.
One must also consider what happens as $\phi_n\rightarrow\infty$. In
the case where $q_n<0$, the theory can have a stable ground state (with
broken \susy) only if ${\kahler ij}^{-1}$ grows sufficiently fast.  We
expect this to be the case if the original theory has no classical flat
directions. Otherwise the theory has no ground state.   \medskip

We believe these arguments explain why previously discovered calculable
models which are known to dynamically break \susy\ also have an R
symmetry. Our arguments that an R symmetry is
necessary for DSB do not apply to noncalculable models, for which there
is no separation between the \susy\ breaking scale and the scale of the
strong gauge dynamics. For instance there are several models which are
believed to dynamically break \susy\ in a regime where the theory is a
strongly coupled chiral gauge theory. Examples include an $SU(2n+1)$
gauge theory, with $n>1$, coupled to chiral superfields in one
antisymmetric tensor and $2n-3$ antifundamental representations, with the
most general renormalizable gauge invariant superpotential consistent
with the symmetries.  Another example is an $SO(10)$ gauge theory with a
single $16$. However in all of these theories there is an accidental R
symmetry, and the criteria of reference \adstwo\ are sufficient to show
that \susy\ is broken.

Our arguments also do not apply to nongeneric models. Below we give
various examples of generic superpotentials to which our arguments do
apply, and of nongeneric superpotentials which are counterexamples.  All
of these should be assumed to have the conventional $\sum_i z_iz_i^*$
form for the Kahler potential.
\lfm{1.}{\it Generic superpotential, exact unbroken R symmetry,
spontaneously broken non-R symmetry, unbroken \susy.} The superpotential
is
\eqn\sxsq{\lambda S X^2\ ,}
which is the most general consistent with the R symmetry under which $S$
has charge zero and $X$ has charge 1, and with a non-R $U(1)$ symmetry
under which $S$ has charge 2 and $X$ has charge $-1$. \Susy\ is unbroken
along the flat direction labeled by $a$
\eqn\vacii{X=0,\quad S=a\ ,}
which leaves the R symmetry unbroken but does break the other $U(1)$
symmetry (we could have considered a broken R symmetry which is a linear
combination of the two $U(1)$s). Note that the Goldstone boson {}from
the spontaneously broken $U(1)$ has a massless scalar partner as
required by \susy, and there is a noncompact flat direction.
\lfm{2.}{\it Nongeneric superpotential, exact spontaneously broken R
symmetry, unbroken \susy.}
This model has superpotential
\eqn\rwsusy{S X^2 +SXY + X^2 \bar Y +mY \bar Y \ ,}
where $S, \bar Y$ are superfields with R charge 2, and $X, Y$ have R
charge 0.  The supersymmetric vacua are
\eqn\vacii{X=Y=\bar Y=0,\quad S=a}
and
\eqn\vaciia{X=m , \quad S=\bar Y=0, \quad Y=-m\ ,}
where $a$ is an arbitrary parameter characterizing a flat direction
associated with the broken R symmetry. Note that this flat direction is
required for supersymmetry, so that the Goldstone boson {}from the
spontaneously broken R symmetry can have a massless scalar partner, and
that the existence of such a flat direction requires a superpotential
such as \rwsusy\ which is carefully chosen so that equations \nmo\ and
\nth\ can be satisfied simultaneously. If other invariant terms such as
$S$ are added, there is only an R preserving supersymmetric ground state
with $S=0$.
\lfm{3.}{\it Generic superpotential, exact R symmetry,
broken \susy.}
This is the O'Raifeartaigh model with the superpotential
\eqn\orsu{\mu^2 S + SQ^2 + mPQ \ ,} which is the most general
renormalizable superpotential consistent with an R symmetry and a $Z_2$
symmetry. All ground states break \susy.  The ground state with $S=0$ does
not spontaneously break the R symmetry. The R symmetry guarantees
that the addition of generic nonrenormalizable terms consistent with the
symmetries will not restore \susy.
\lfm{4.}{\it Nongeneric superpotential, no R
symmetry, broken \susy.} The superpotential \eqn\ormod{\lambda_x
XQ^2+\lambda_y Y(Q^2-\mu^2)+\lambda Q^3+m_1 Q^2 +m_2^2 Q} has no R
symmetry, but no supersymmetric vacua.  The addition of other terms
nonlinear in $X$ or $Y$, which are allowed by all symmetries, would
restore \susy.

\medskip

We have argued that in flat space any generic calculable model has DSB
if it has a spontaneously broken R symmetry (an exact R symmetry being
also a necessary condition).  However, when the effects of {\it local}
\susy\ are included, the F-term conditions are replaced by
\eqn\sugracond{D_j W = W_j+{1\over M_P^2}K_j  W =0\quad ({\rm
supergravity})\ .}
In order to ensure that the effective cosmological constant is zero
after \susy\ is broken, $W$ must be of order $M_s^2 M_P$, where $M_s$ is
the \susy\ breaking scale.  For a theory which exhibits \susy\ breaking
in flat space, since $W_j\sim M_s^2$ and $K_j=\CO (\vev z )$, the
additional terms in \sugracond\ are negligible compared to $W_j$ unless
some field strengths are of order $M_P$, in which case any effective
field theory description breaks down and we cannot answer the question
of whether or not there is a supersymmetric minimum without a detailed
understanding of Planck scale physics.  We thus believe that our
analysis is the most general one possible within the regime of validity
of effective field theories.

This discussion has an immediate application to some popular models of
supersymmetry breaking in string theory.  The original version of these
models
\ref\glucon{J.P. Derendinger, L.E. Ibanez and H.P. Nilles,
\pl{155}{1985}{65}; M. Dine, R. Rohm, N. Seiberg and E. Witten,
\pl{156}{1985}{55}}
was based on gluino condensation in some gauge group.  A dimension five
operator
\eqn\gluor{\int d^2 \theta S W_\alpha^2}
couples the gauge superfields $W_\alpha$ to a dilaton field $S$ and
induces upon integrating out the gauge dynamics an effective
superpotential for $S$
\eqn\firsteff{W_{\rm eff} = e^{-CS} }
with some constant $C$ depending on the gauge group.  Both the original
theory with the coupling \gluor\ and the effective theory with the
superpotential \firsteff\ exhibit an anomaly free R symmetry under
which $S$ is shifted by an imaginary amount.  As follows {}from our
general analysis supersymmetry is broken for any finite $S$ where the R
symmetry is broken.  The problem with this model is that the potential
for $S$ slopes to zero at infinity and the theory does not have a ground
state.

In order to avoid this problem, the authors of reference
\ref\racetra{N.V. Krasnikov, \pl{193}{1987}{37}; L. Dixon, V.
Kaplunovsky, J. Louis and M. Peskin, unpublished;  L. Dixon, talk
presented at the APS, DPF Meeting at Houston (1990); V. Kaplunovsky,
talk presented at the ``String 90'' workshop at College Station (1990)}
proposed to add another gauge group and to couple it to $S$ as in
\gluor.  These models are referred to as ``race track models.'' The
effective superpotential obtained after integrating out the two gauge
sectors is
\eqn\seceff{W_{\rm eff} = e^{-C_1 S} + A e^{-C_2 S}}
where the constants $C_i$ depend on the two gauge groups.  Unlike the
previous situation, now there is no exact R symmetry.  Correspondingly,
the superpotential \seceff\ leads to a supersymmetric ground state at
finite $S$.  More sophisticated versions of this proposal
\ref\moreso{See \eg: S. Ferrara, N. Magnoli, T.R. Taylor and G.
Veneziano, \pl{245}{1990}{409}; J.A. Casas, Z. Lalak, C. Munoz and G.G.
Ross, \np{347}{1990}{243}; B. deCarlos, J.A. Casas and C. Munoz,
\pl{263}{1991}{248}}
involve also some other fields (moduli).  Since more than one gauge
group is used, the models do not have an R symmetry and therefore, as
follows from our general analysis, they all exhibit supersymmetric
ground states at finite field strength (as well as some
nonsupersymmetric states).

\newsec{Explicit R symmetry breaking and the R-axion}

The necessity of a spontaneously broken R symmetry for \susy\ breaking
is troublesome, since it would imply the existence of a Goldstone boson,
the R-axion, which could be in conflict with phenomenological and
astrophysical observations.  Also we do not expect on theoretical
grounds that such a global symmetry can be exact \refs{\worm,\ \noglob}.
Fortunately, we have found that several ways of explicitly breaking the
R symmetry can be shown not to restore supersymmetry. This is a
consequence of the lack of genericity of many supersymmetric models and
of a nonperturbative nonrenormalization theorem \nonren.

\subsec{Higher dimension operators}

There are many examples of models with an accidental R symmetry of the
gauge invariant renormalizable terms, but in which this symmetry could
be explicitly broken by higher dimension operators in either the
superpotential or the Kahler potential.  The following argument shows
that for a general class of theories such symmetry breaking can only
lead to \susy\ restoration for field strengths which are at least as
large as the dimensionful scale which suppresses the effects of the
higher dimension terms.

\def\ls{{\Lambda_S}}
\def\fs{{\Lambda_S^2}}
\def\wdy{{W_{\rm dyn}}}
\def\wr{{W_{\rm ren}}}
\def\we{{W_{\rm eff}}}
\def\wn{{W_{\rm nonren}}}
\def\ftermi{{\partial\we\over \partial z_i}}
\def\fdi{{\partial \wdy\over \partial z_i}}
\def\wt{{W_{\rm tree}}}
\def\fti{{\partial \wt\over \partial z_i}}

Consider a theory of DSB with effective superpotential $\we=\wr+\wdy,$
with $\wr$ a renormalizable superpotential only containing terms of
dimension four, and $\wdy$ a term generated by the nonperturbative
dynamics of an asymptotically free gauge theory.  This gauge theory
generates a scale $\ls$ through dimensional transmutation.  On general
grounds $\wdy$ must be proportional to a positive power of $\ls$. (The
restriction that $\wdy$ can only contain soft terms is an example of the
lack of genericity of supersymmetric effective superpotentials.)  We
will work in a canonical basis for the fields, where the Kahler
potential is $\sum_i z_iz_i^* +$ higher dimension terms. Now we also
assume

\lfm{1.} $\we=\wr+\wdy$ has an exact  continuous $U(1)$ R symmetry.
\lfm{2.} The only dimensionful terms in $\we$ are characterized by the
scale $\ls$.
\lfm{3.} This theory spontaneously breaks \susy\ (and R symmetry) at the
scale $\ls$. Therefore, for all values of the fields
\eqn\hdi{\sum_i \abs{\ftermi }^2+\sum_A\(z^*_j
T^A_{jk}z_k\)^2\ge\CO(\ls^4)\ .}
\lfm{4.} The theory without $\wdy$ has no flat directions, and the tree
level theory is scale invariant. Thus since $\wdy$ is generated by an
asymptotically free interaction, it must be proportional to a positive
power of $\ls$ and hence must grow more slowly than $z^3$ for large field
strength. Therefore, for any field strength $z$ which satisfied the D
flatness condition and which is much larger than $\ls$,
\eqn\hdii{\sum_i\abs{\ftermi}^2\sim z^4\ .}
\medskip

Now we wish to study whether higher dimension operators can restore
\susy. We assume these operators are characterized by at least one
inverse power of a scale $m$. Higher dimension terms in the Kahler
potential can help restore \susy\ only if ${\kahler ij}^{-1}$ has
zero eigenvectors, which can happen only for field strengths of order
$m$. We can also add higher dimension terms to $\we$. Call the higher
dimension terms $\wn$, which can be written
\eqn\hdiii{\wn=m^3 f(z_i/m)\ ,\quad f(x)\le x^4\ {\rm when}\ x<1\ .}
These terms can only restore \susy\ for
\eqn\hdiv{{\partial\wn\over \partial z_i}\ge\fs\ ,}
which can only happen for field strengths
\eqn\hdv{z\ge (m\fs)^{1/3}\ ,}
which is greater than $\ls$. However for such large field strengths,
because of our assumption 4,
\eqn\hdvi{\ftermi\sim z^2\ .}
These F terms can only be canceled out by contributions {}from $\wn$
when the light field strengths are of order $m$.

We therefore conclude that if we have a renormalizable theory with DSB
and no classical flat directions, higher dimension operators can only
restore \susy\ for field strengths of order the scale which
characterizes them -- typically the Planck mass.  Understanding whether
or not this actually happens requires a complete understanding of Planck
scale physics.

If the higher dimension operators arise {}from integrating out heavy
particles with a mass less than the Planck mass, then we can study the
full theory to understand whether or not \susy\ is actually restored at
large field strengths.  In many cases there will not be any
supersymmetric minimum, as shown by the following argument.

Consider a renormalizable theory which, in the limit where some massive
parameter $m$ goes to infinity, has a renormalizable low energy
effective theory with an unbroken asymptotically free gauge theory, a
continuous R symmetry, no flat directions, spontaneously broken \susy,
and no dimensionful parameters other than $\ls$, the \susy\ breaking
scale in the effective theory.  $\ls$ will generally have some $m$
dependence but we assume that there is a gauge parameter $g$ which we
can tune to make $\ls(g,m)\ll m$.  We want to understand whether the
theory has a supersymmetric minimum. By our previous arguments, if such
a minimum does exist, then when $m$ is larger than $\ls$, at least some
of the light fields must have vevs of order $m$ or larger at the
minimum.  Now we also assume that the full theory has no classical flat
directions, and that classically when any of the fields have vevs of
order $m$ that there is at least one F or D term which is as large as
$\CO(m^2)$). Let us only consider D flat directions. The full
superpotential is $\we=\wt+\wdy$, where $\wdy$ is characterized by a scale
$\ls$ which can be made arbitrarily small compared with $m$. A
supersymmetric minimum requires that for all superfields
\eqn\hdvii{\fti+\fdi=0\ .}
If there is a supersymmetric minimum we know that at least one of the
light fields in the low energy effective theory has vev larger than
$\CO(m)$ and therefore there is at least one field for which
\eqn\hdix{\abs{\fti}\ge\CO(m^2)\ .}
So for this field we must have
\eqn\hdx{\abs{ \fdi }=\abs{\fti}\ge\CO(m^2)  \ .}
Thus a supersymmetric minimum
could only exist for field strengths larger than $m$ and then only if
there is a contribution to $\wdy$ which for large field strength grows in
some direction in the light field space faster than any of the tree level
terms. Such a situation is possible when there are directions involving
the light fields for which the classical potential grows more slowly than
quartically. However there are many models, such as the example we will
present in the next section, for which the potential grows quartically
for any direction in the  light  field space, and which therefore
cannot possess a supersymmetric minimum at any field strength.

This argument has assumed that $m\gg\ls$. However in a supersymmetric
theory it is not possible to have \susy\ unbroken for a range of
parameters but broken for other values \wittentwo. The only exception
to this rule is when the behavior of the theory at large field strengths
changes \wittentwo.  We conclude that in such a  model there is no
finite  range of $m$ for which \susy\ is unbroken.

If one can find a calculable model where \susy\ is dynamically broken,
even though there is no R symmetry, then $\we$ cannot be generic.  In
the next section we study an example of such a model, and show how the
lack of genericity can be exploited to give DSB.

\subsec{A model of dynamical supersymmetry breaking without an R symmetry}

The simplest known calculable model of DSB, which was analyzed in detail
in reference \adstwo, has gauge group $SU(3)\times SU(2)$, and
superfields
\eqn\modi{Q\sim(3,2),\quad  \bar{U}\sim(\bar3,1), \quad
\bar{D}\sim(\bar3,1),\ {\rm and}\quad L\sim (1,2)\ . }
The classical superpotential is
\eqn\modii{ W=\lambda  Q\bar{D} L\ .}
This theory has no flat directions and an exact spontaneously broken R
symmetry.  The $SU(3)$ gauge theory gets strong at a scale $\ls$ which
is higher than the $SU(2)$ scale and spontaneously breaks \susy\ at
a scale $\lambda^{5/14}\ls$. There is also an exact global $U(1)$
symmetry, called hypercharge, which is not spontaneously broken.

In order to study the effects of explicit R symmetry breaking, we add to
this theory two chiral superfields $S$ and $\bar S$ which are $SU(2)$
singlets and transform like a triplet and an anti-triplet under $SU(3)$.
The hypercharge assignments of these fields are
\eqn\modiii{Q\sim\frac16,\quad\bar{U}\sim-\frac23,
\quad\bar{D}\sim\frac13,\quad  L\sim-\frac12,\quad S\sim-\frac13,\quad
\bar{S}\sim\frac13\ . }
The most general renormalizable tree level superpotential invariant
under hypercharge is
\eqn\trees{W_t= \lambda L Q \bar D + \lambda_s L Q \bar S +\lambda' Q^2
S + \bar \lambda\;\bar D\bar U\bar S + m \bar S S.}
When $m$ is large, we can integrate out $\bar S$ and $S$ and find the
effective tree level superpotential for the light fields
\eqn\ntree{W_N= \lambda L Q \bar D + {\lambda' \bar \lambda \over m}Q^2
\bar D \bar U .}
For $\lambda'$ or $\bar \lambda =0$ the model has an R symmetry and no
flat directions and therefore satisfies the conditions of reference
\adstwo\ for DSB. When $\lambda',\bar \lambda \not= 0$ but $m=0$ there
is no R symmetry and there is a classical flat direction with nonzero $S$
and $\bar S$.

We wish to know whether \susy\ is broken in this theory when $m$ and all
the $\lambda$'s are nonzero.  Naively, the answer is yes, since when $m$
is large, $S$ and $\bar S$ decouple and we are left with a theory which
is known to break \susy.  Then by the arguments of reference \wittentwo,
\susy\ should be broken for any value of $m$. This model is a specific
example of the general class of theories described in the previous
subsection, which we argued would have to break \susy\ for large $m$
unless there is a direction in the light field space for which
dynamically generated terms grow faster than the tree level potential.
The only possible such direction is large $S$ and $\bar S$, since in
this direction the potential only grows quadratically. However these
fields decouple in the large $m$ limit and the supersymmetric minimum
must involve expectation values larger than $m$ for at least one of the
light fields $Q$, $\bar U$, $\bar D$, and $L$, and in all such
directions the classical potential grows quartically. Thus this model
cannot have a supersymmetric minimum, even though it has no R symmetry,
due to the lack of genericity of the effective superpotential.


Let us analyze the low energy effective theory given by \ntree\ in more
detail.  The scale at which the low energy $SU(3)$ theory becomes strong
is $\ls= m^{1/7} \Lambda_3^{6/7}$, where $\Lambda_3$ is the $SU(3)$
scale when $m=0$. If we ignore the small nonrenormalizable term in
\ntree, and the term proportional to $\lambda$, then the effective
theory is massless QCD with two flavors, for which one can show that
instantons generate the effective superpotential
\eqn\wins{W_{\rm instanton} = {\ls^7 \over Q^2 \bar U \bar D} \ ,}
(the $SU(3) \times SU(2)$ gauge indices in $Q^2 \bar U \bar D$ are
contracted in a gauge invariant way) which is the only term allowed by
the symmetries.  Adding the tree level terms we find
\eqn\weffec{\weff = \lambda L Q \bar D + {\lambda' \bar \lambda \over
m}Q^2 \bar U \bar D + {m \Lambda_3^6 \over Q^2 \bar U \bar D}\ .}
The R invariance of $\weff$ is broken by the nonrenormalizable term.
However this effective superpotential does not lead to a supersymmetric
minimum.  To see this, note that if the equation $\partial_{L_i}W_{eff}
= \lambda Q_i \bar D=0$ is satisfied (here $i$ is the $SU(2)$ index),
and the D-term equations for the $SU(3)$ and $SU(2)$ are also satisfied,
then $Q^2 \bar U \bar D=0$ and the superpotential is singular.

To compare with our analysis in section~2, the superpotential \weffec\
is not generic.  The second term breaks the R symmetry but does not
restore \susy. However we have so far neglected the possibility that
nonzero $\lambda$, $\lambda'$ and $\bar \lambda$ could generate
additional nonperturbative terms, such as $\ls^{7/2}\left( (LQ
\bar D) / (Q^2 \bar U \bar D) \right)^{1/2}$ which would restore
\susy.  However the nonperturbative nonrenormalization theorem of
reference \nonren\ shows that this does not happen, the effective
superpotential \weffec\ is exact and does not receive any corrections.
In other words, {\it the superpotential is not generic and \susy\ is
broken.}

\subsec{Effects of higher dimension operators on the R axion mass}

Despite the example in the previous section, it remains true that in all
known models of DSB which have no mass scales in the classical
Lagrangian, there is a massless R axion. In realistic models, the R
axion gains a small mass {}from the QCD anomaly.  This axion could be in
conflict with astrophysical and phenomenological observations if the
scale of R symmetry breaking is lower than $\sim 10^{10}$ GeV and with
cosmology if the scale is higher than $\sim 10^{12}$ GeV
\nref\invisaxion{Relevant reviews include J.E.  Kim,
\prep{149}{1987}{1};
R. Peccei, published in {\it CP Violation}, ed. C. Jarlskog (1988)}
\nref\bkn{T. Banks, D. B. Kaplan, and A. E. Nelson, preprint
UCSD/PTH 93-26, RU-37 (1993)}
\refs{\invisaxion,\ \bkn}. However we now know that the R symmetry could
be explicitly broken by nonrenormalizable terms without restoring \susy.
Such terms could solve the astrophysical problems associated with the
R axion occurring in models of DSB at low energy
\ref\dynsusy{M. Dine and A.E. Nelson, \physrev{48}{1993}{1277}},
by giving the axion a mass which is too large to be produced in stars.

If the R symmetry is explicitly broken by operators of dimension five
proportional to $1/m_P$, then the effects on the R axion mass are quite
interesting. Assuming that the only dimensionful scale in the low energy
effective theory is the \susy\ breaking scale $\ls$, and that this is
also the scale of spontaneous R symmetry breaking, the R axion mass
squared is approximately
\eqn\axionmass{m_{\rm axion}^2\sim\ls^3/m_P\ .}
For $\ls$ of order $10^5$ GeV, this gives $m_{\rm axion}\sim 10$ MeV,
which is just barely light enough to be produced in supernovae. We
conclude that the R axion is not an astrophysical problem for models of
DSB in the visible sector, provided that the R symmetry can be broken by
terms of dimension 5 and $\ls>10^5$ GeV.

There can also be cosmological problems associated with overproduction
of the R axion in the early universe, if the R symmetry breaking scale
is higher than $\sim 10^{12}$ GeV
\ref\cosmoaxion{J. Preskill, M.B. Wise, and F. Wilczek,
\pl{120}{1983}{127}; L. P. Abbott and P. Sikivie, \pl{120}{1983}{133};
M. Dine and W. Fischler, \pl{120}{1983}{137}}.
A general analysis of the cosmological bounds on the axion coupling and
mass
\ref\enq{J.  Ellis, D.V. Nanopoulos and M. Quiros, \pl{174}{1986}{176}}
shows that this problem can also be solved if explicit symmetry breaking
gives the axion a large enough mass. For instance an  axion with a mass
given by eq.~\axionmass\ does not cause any cosmological problems when
the scales  of R symmetry breaking and \susy\ breaking are the same.
However in models where dynamical \susy\ breaking is driven by
nonrenormalizable operators suppressed by the Planck mass, the R axion
mass is typically not large enough to avoid cosmological overproduction
\bkn.

\subsec{R color}

Another method for solving the R axion problem \adstwo, is to introduce
a new strong gauge group called R color, under which the R symmetry has
an anomaly, in addition to the group whose dynamics are responsible for
DSB.  Nonperturbative R color effects can give the R axion a mass, but
one must then worry about whether they can also induce new terms in the
effective superpotential which could restore \susy. The following
argument shows that in many cases R color does not restore \susy, due to
the lack of genericity of the nonperturbatively generated terms.

\def\lr{{\Lambda_R}}
\def\fsi{{\partial W_S\over \partial z_i}}
\def\fri{{\partial W_R\over \partial z_i}}

Consider a theory whose classical lagrangian only has dimension four
terms, with no classical flat directions, a spontaneously broken R
symmetry, and supersymmetry breaking scale $\ls$. Now introduce R color
by gauging a global symmetry of this theory, which becomes strong at a
scale $\lr\ll\ls$.  R color can nonperturbatively induce operators in
the effective superpotential which always grow more slowly for large
field strength than $\CO(z^3)$, and which break the R symmetry.

We write the effective superpotential as
\eqn\rci{\we=W_S+W_R\ ,}
where $W_S$ is the effective superpotential of the theory in the limit
where R color is turned off, and the remaining terms in $W_R$ are
generated by nonperturbative R color effects. For all D-flat directions
there is at least one superfield for which
\eqn\rcii{\abs{\fsi}\ge\fs \ .}
\Susy\ restoration requires that
\eqn\rciii{\fsi+\fri=0\ ,}
thus to restore \susy\ it is necessary that
\eqn\rciv{\abs{\fri}\ge\fs\ .}
R color can only induce terms proportional to positive powers of $\lr$,
and for field strengths larger than $\CO(\ls)$
\eqn\rcv{\abs{\fri}<\CO(z^2) \ .}
Because of the lack of classical flat directions, for such field
strengths
\eqn\rcvi{\abs{\fsi}\sim \CO(z^2)\ ,}
and so it is not possible to satisfy eq.~\rciii\ for field strengths
larger than $\CO(\ls)$.  So  R color could only restore \susy\ in the
region where all field strengths are smaller than $\ls$ and the \susy\
breaking group is strongly coupled.  In this region  the contribution of
the \susy\ breaking group to the effective potential is larger than
$\CO(\ls^4)$, and  R color would have to make an equally important
canceling contribution in order to restore \susy. This is not possible
since we are assuming that $\lr\ll\ls$, and the R color contribution to
the effective potential will be smaller than
$\max\(\CO(\lr^4),\CO(z^4)\)$.

We conclude that when R color is added to a DSB model with no classical
flat directions or mass scales, then if R color is sufficiently weak
nonperturbative R color effects cannot restore \susy. Furthermore, we
have previously noted that it is not possible to have a theory in which
\susy\ is unbroken for a range of parameters but broken for other
values, so \susy\ breaking should persist even when R color is as strong
or stronger than the \susy\ breaking gauge group.

\newsec{Conclusions}

We have shown that a continuous R symmetry is a necessary condition for
spontaneous \susy\ breaking and a spontaneously broken R symmetry is a
sufficient condition, in models where the gauge dynamics can be
integrated out and in which the effective superpotential is a generic
function consistent with the symmetries of the theory. Therefore the
existence of a nonanomalous R symmetry is a useful guideline for
finding a model with dynamical supersymmetry breaking. This implies that
a troublesome Goldstone boson, the R axion, is a typical feature of
models which break \susy\ dynamically.

We have also argued that because supersymmetry severely limits the terms
which can be generated nonperturbatively in the effective
superpotential, in some cases the effective superpotential is {\it not}
generic and it is possible to dynamically break \susy\ {\it without} an
exact R symmetry.  The R axion can be given an acceptably large mass,
either by introducing higher dimension operators which explicitly break
the R symmetry or by introducing a new gauge group under which the R
symmetry is anomalous. We conclude by speculating on the attractive
possibility that because of the lack of genericity of effective
superpotentials, there may exist a model with a single gauge group, in
which all mass scales arise through dimensional transmutation, \susy\ is
spontaneously broken, and there is no R symmetry or R axion at all.
Such a model would be an excellent candidate for dynamical \susy\
breaking near the weak scale.

\centerline{\bf Acknowledgments}

It is a pleasure to thank T. Banks, D. Kaplan, S. Shenker and E. Witten
for several useful discussions. A.N. would like to thank Rutgers
University and the CERN theory group for their hospitality. This work
was supported in part by DOE grants \#DE-FG05-90ER40559 and \doe, and by
the Texas National Laboratory Research Commission under grant \tnlrc.
A.N. was supported in part by an SSC fellowship and a Sloan fellowship.

\listrefs
\bye